\documentclass[11pt]{article}
\usepackage{times}\usepackage{mathptmx}
\newcommand{\atitle}[1]{{\centering\noindent\bf\hspace*{0.5cm} #1\par}\vspace{0.2cm}}
\newcommand{\auth}[1]{{\centering\noindent\hspace*{0.5cm} #1\par}\vspace{0.3cm}}
\newcommand{\inst}[1]{{\centering\parbox[t]{0.8\textwidth}
    {\centering\it\noindent#1\par}\par}\vspace{0.6cm}}
\newcommand{\emailadd}[1]{{\centering\parbox[t]{0.8\textwidth}
    {\centering\it\noindent#1\par}\par}\vspace{0.6cm}}
\newcommand{\txt}[1]{\vspace{0.1cm}#1\par}
\newcommand{\refs}[1]{\vspace{0.11cm}\noindent
\begin{list}{[1]}{\parsep0pt\itemsep0pt\topsep0pt\partopsep0pt%
\labelsep0.1cm}#1\end{list}}
\begin{document}
Abstract for "Molecules in Space \& Laboratory", Paris, France\\
 May 14--18, 2007
\par\vspace{3cm}%
\setlength{\unitlength}{1cm}%
%
%
%
\atitle{%
The end of the "Empty Field" epoch in optical identifications
}
\auth{%
Lipovka A.,$^{1}$
Lipovka N.,$^{2}$
}
\inst{%
$^{1}$ Department for Physical Research, University of Sonora, Rosales y Blvd. Transversal, col. Centro, edif. 3-I, Hermosillo, Sonora, 83000, Mexico\\
$^{2}$ St.Petersburg branch of Special Astrophysical Observatory, Pulkovskoye sh. 65. St.Petersburg, 196140, Russia\\
}
\emailadd{%
aal@cajeme.cifus.uson.mx
}


\txt{%
In order to obtain more comprehensive information about an celestial
object, the radio image must be identified with the optical one.
Many years the identification process is carried out with the
coordinate coincidence criteria, which leads to abundant
misidentifications and "empty field" in optics for the radio
sources. For this reason significant part of radio sources do not
have identifications in optic. In present paper we consider the
radio refraction in the Galaxy, which significantly changes the
coordinates of radio sources if compared with the optical one. By
taking into account the radio refraction, the major number of the
radio sources can be successfully identified with the optical
objects. For our calculation of the radio refraction we use the ISM
model discussed in [1,2,3,4] and [5]. The coordinate correction for
the refraction at 21 cm wavelength (NVSS) consists typically some
arcminutes for the distant galaxies, but in some particular cases
can reach tenth of arcmin. The method of the optical identifications
corrected for the refraction is developed for the Galaxy ISM. To
illustrate, it was applied for some NVSS maps which cover more than
10 percents of the sky. The results are presented at Figures 1 and
2. At the Fig.1 one can see the distribution of the galaxies as
function of the coordinate correction for the refraction. Fig.2
shows the calculated radio refraction for the ISM model, as function
of the frequency. The upper bold line corresponds to the maximum
value of refraction for the used model, and lower fine line to the
mean (more probable to observe) refraction. By open circles at same
figure we mark the observed refraction obtained for some radio
sources of the NVSS maps (21 cm). One can see an excellent agreement
between observed and calculated refraction.

\begin{figure}[tbp]
\vspace{7cm} \includegraphics{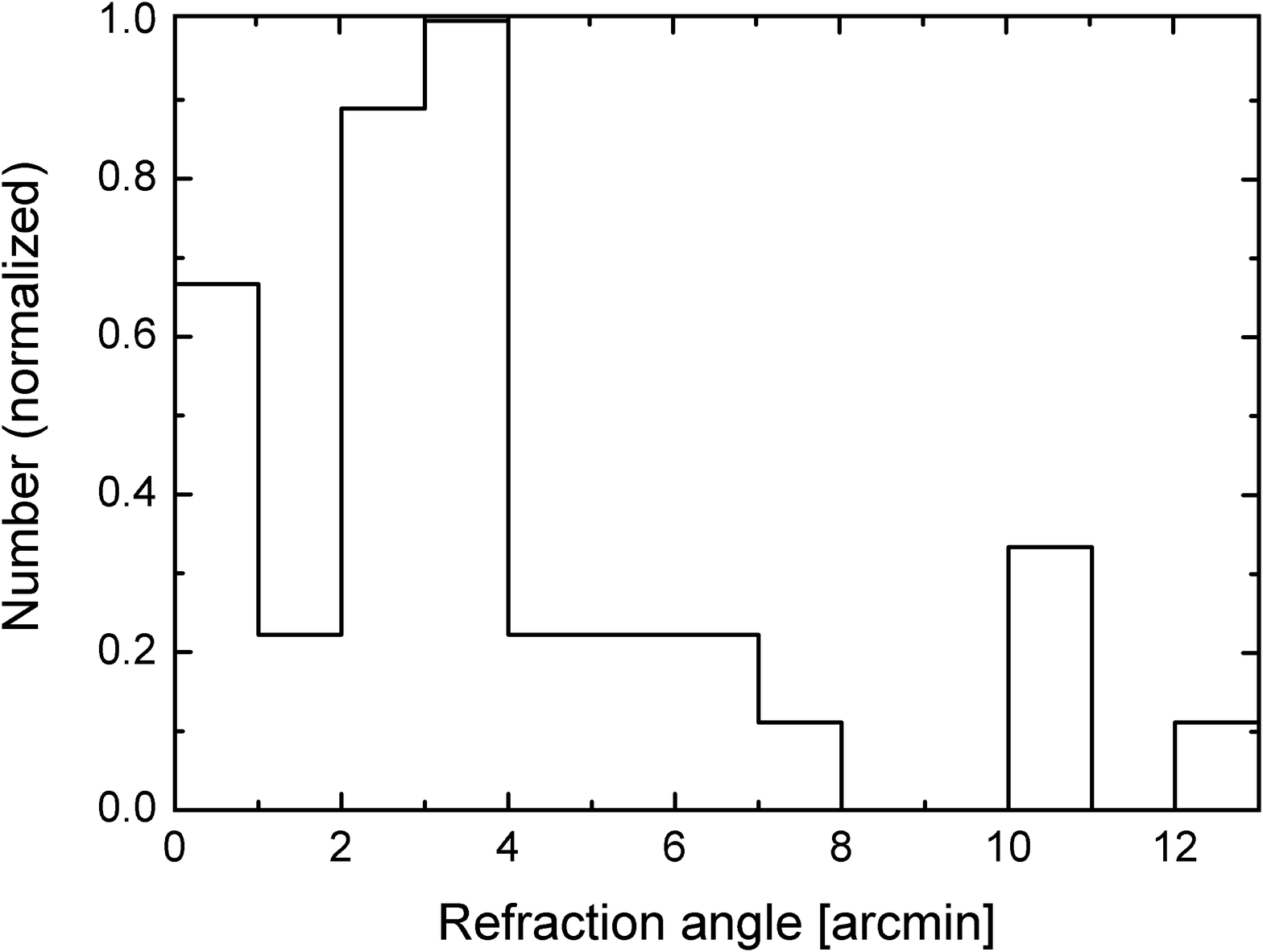} \caption{}
\end{figure}

\begin{figure}[tbp]
\vspace{7cm} \includegraphics{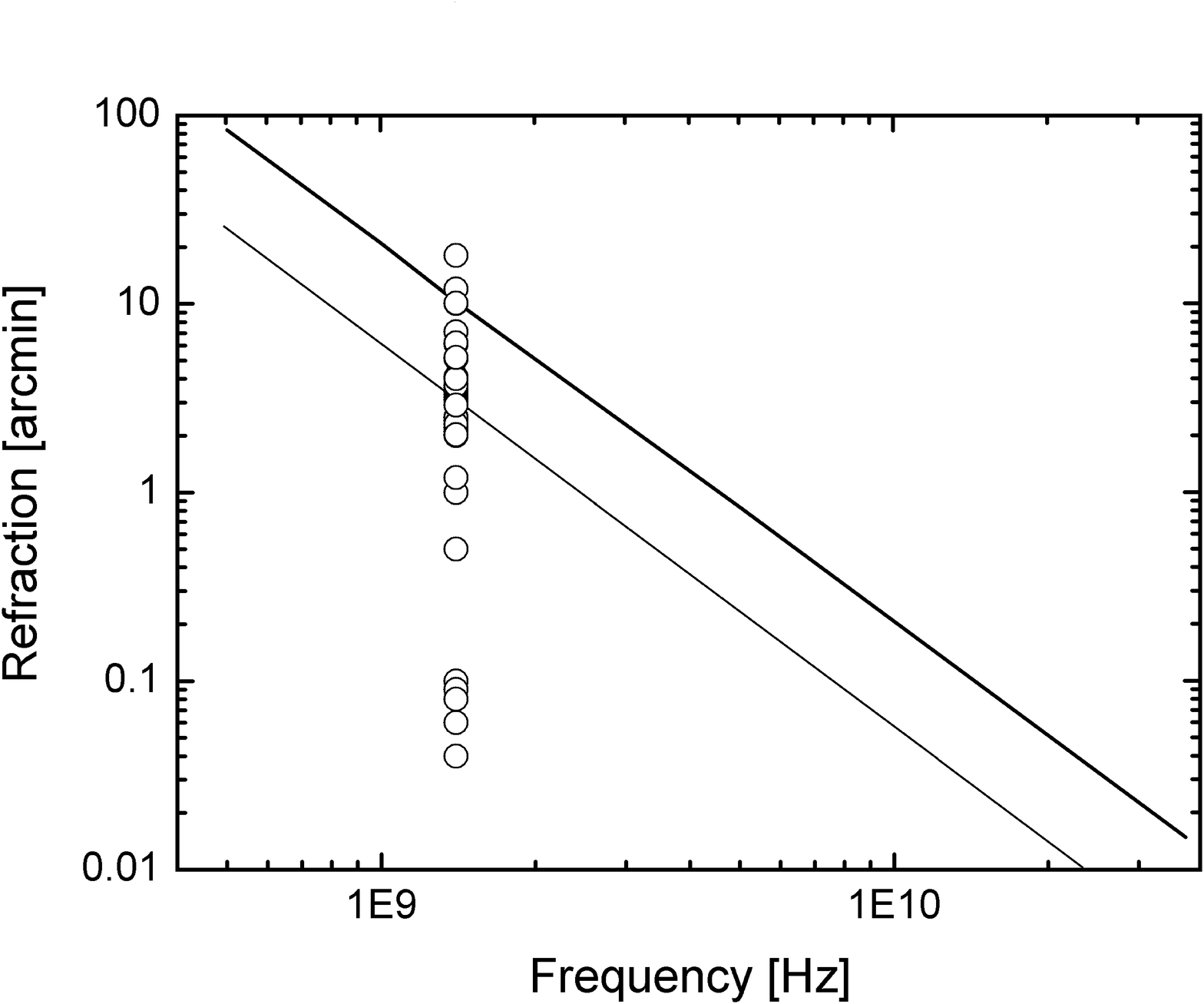} \caption{}
\end{figure}

Discovered here first time radio refraction is an powerful tool for
investigation of the physical conditions in ISM and can also be used
for independent measurements of the distances to the galaxies.

}
\refs{%
\item[{[1]}] N.M. Lipovka 1977, Sov. Astron. 21, 151
\item[{[2]}] A.P. Venger, I.V. Gosachinskii, V.G. Grachev, T.M. Egorova, N.F. Ryzhkov, V.K. Khersonsky 1984, Aph.S.S., 107, 271
\item[{[3]}] I.V. Gosachinskii, V.K. Khersonsky 1984, Aph.S.S., 107, 289
\item[{[4]}] I.V. Gosachinskii, V.K. Khersonsky 1985, Aph.S.S., 108, 303
\item[{[5]}] K. Wada, C. Norman 2007, (ApJ accepted) arXiv:astro-ph/0701595
}
\end{document}